\documentclass[twocolumn,aps,a4paper,showpacs,amsfonts,amsmath,amssymb]{revtex4}
\usepackage{makeidx}
\usepackage{graphicx}
\usepackage{epsfig}

\begin{document}

\title{Extreme values and the level-crossing problem. \\ An application to the Feller process} 

\author{Jaume Masoliver}
\email{jaume.masoliver@ub.edu}
\affiliation{Departament de F\'{\i}sica Fonamental, Universitat de Barcelona,\\
Diagonal, 647, E-08028 Barcelona, Spain}

\date{\today}

\begin{abstract}

We review the question of the extreme values attained by a random process. We relate it to level crossings either to one boundary (first-passage problems) and two boundaries (escape problems). The extremes studied are the maximum, the minimum, the maximum absolute value and the range or span. We specialize in diffusion processes and present detailed results for the Wiener and Feller processes.

\end{abstract}

\maketitle

\section{Introduction}
\label{sec1}

Level-crossing problems --including first-passage and escape problems-- have a long and standing tradition in physics, engineering and natural sciences, with great theoretical interest in, for instance, bistability and phase transitions and countless practical applications ranging from meteorology, seismology, reliable theory, structural and electrical engineering and finance, just to name a few \cite{gardiner,redner1,weiss1,katja,condamin,mporra,west,shlesinger,eichner,salvadori,maso_pere_07,maso_pere_08,maso_pere_09,mont_maso,
mario}.  

The level crossing problem is closely related to the theory of extremes, the latter initiated in the late nineteen twenties by the works of Frechet, Fisher and Tippet and subsequently developed by Gnedenko and Gumbel later in the forties and early fifties \cite{gumbel}. It applied to series of independent random variables and the central result is the Frechet-Tippet theorem which states that under suitable conditions the asymptotic distribution of extremes are restricted to be of three types (Gumbel, Frechet and Weibull) \cite{katja,gumbel,leadbetter}. As remarked in Refs. \cite{katja} and \cite{west}, when extreme events are rare (which is often the case) they can be approximately treated as independent variables for which the Fisher-Tippet theorem holds. This approximation, however, reduces the question to a problem of statistics and time series analysis and neglects the underlying dynamics and the correlations induced by it.    

The extreme-value problem basically includes the maximum and minimum values attained by a given random process during a certain time interval. It also encompasses the range or span defined as the difference between the maximum and the minimum. In physics this problem has been traditionally related to level crossings and first-passage times and it has been basically restricted to diffusion processes \cite{katja,darling,blake} (see also \cite{berman} for similar developments aimed also to diffusion processes but oriented to the pure mathematician).  

This is a complicated business because obtaining first-passage probabilities is essentially difficult. This is one of the reasons why, to my knowledge, few exact analytical approaches have appeared except for the Wiener process and, to a less extend, for the Ornstein-Uhlenbeck process \cite{katja,darling,blake}. Despite the intrinsic difficulty there are, however, recent works investigating this kind of problems in subdiffusions and other anomalous diffusion processes as well (see \cite{yuste} and references therein). 

In a recent paper \cite{mp-feller} we have studied the first-passage problem for the Feller process and presented a complete solution of it, including first-passage and exit probabilities and mean first-passage and mean exit times. One of our goals here is to apply those results to obtain the extreme values attained by the Feller process. Another objective is to review the link between level crossings and extremes by presenting a complete account of the  results  involved (some of them in a new and simpler form) because the connection among both problems is not widely known in the current physics literature. 

In level-crossing problems the issue of primary interest is to ascertain the statistical information on the time taken by a random process to reach, or return to, a given boundary for the first time. If the boundary consists of only one point --which we usually call critical value or threshold-- one deals with a first-passage or hitting problem. If the boundary consists of two points we have an escape or exit problem out of the interval spanned by the boundary points.  As we will see maximum and minimum are the extremes related to the hitting problem while the maximum absolute value and the span are related to the exit problem.  

The paper is organized as follows. In Sec. \ref{sec2} we review the relationship between first-passage and extreme-value problems. In Sec. \ref{sec3} we review the link between  the escape problem and, both, the maximum absolute value and the span. In Secs. \ref{sec4} and \ref{sec5} we explicitly obtain these results for the Wiener and Feller processes respectively. A short summary of main results is presented in the last section. Some mathematical proofs and more technical details are in appendices.

\section{First passage and extremes}
\label{sec2}

The hitting problem of a random process $X(t)$ is solved if we know the first-passage probability, $W_c(t|x)$, of reaching for the first time threshold $x_c$ when the process starts at $x=X(t_0)$ at some initial time $t_0$ (in what follows we deal with time-homogeneous processes so that $t_0=0$). In terms of the hitting probability the survival probability --{\it i.e.}, the probability $S_c(t|x)$ that at time $t$, or during any previous time, the process has not reached $x_c$-- is simply given by
\begin{equation}
S_c(t|x)=1-W_c(t|x).
\label{s_w}
\end{equation}

For one-dimensional diffusion processes charaterized by drift $f(x)$ and diffusion coefficient $D(x)$, the hitting probability satisfies the Fokker-Planck equation (FPE) \cite{gardiner,mp-feller}
\begin{equation}
\partial_t W_c(t|x)=f(x)\partial_x W_c(t|x)+\frac 12 D(x)\partial^2_{xx} W_c(t|x),
\label{fpe}
\end{equation}
with initial and boundary conditions given by
\begin{equation}
W_c(0|x)=0, \qquad   W_c(t|x_c)=1.
\label{initial}
\end{equation}
Equation (\ref{s_w}) shows that the survival probability obeys the same FPE but with initial and boundary conditions reversed.    

We will now relate the first-passage problem with the extreme values (the maximum and the minimum) reached by the process during a given interval of time. There are other extremes, such as the range or span, which will be discussed in the next section.  

\subsection{The maximum}

We denote by $M(t)$ the maximum value reached by $X(t)$ over the time span $(0,t)$. Formally,
$$
M(t)=\max\{X(\tau); 0\leq\tau\leq t\}.
$$
Note that $M(t)$ is a random quantity whose value depends on the particular trajectory of $X(t)$ and its distribution function is defined by
\begin{equation}
\Phi_{\rm max}(\xi,t|x)= {\rm Prob}\{M(t)<\xi|X(0)=x\}. 
\label{F_max_def}
\end{equation}
In order to relate this function with the hitting probability we distinguish two cases: $\xi>x$ and 
$\xi<x$. Suppose first that the value of the maximum $\xi$ is greater than the initial value, $\xi>x$, in this case the process $X(t)$ has not crossed threshold $\xi$ at time $t$ and the probability of the event $\{M(t)<\xi|X(0)=x\}$ equals the survival probability $S_\xi(t|x)$. That is
$$
\Phi_{\rm max}(\xi,t|x)=S_\xi(t|x), \qquad (\xi>x).
$$
If on the other hand the value of the maximum is lower than the initial point, $\xi<x$, the event $\{M(t)<\xi|X(0)=x\}$ is impossible and has zero probability. In other words, $\Phi_{\rm max}(\xi,t|x)=0$, if $ \xi<x$. We summarize both cases into the single expression:
\begin{equation}
\Phi_{\rm max}(\xi,t|x)=S_\xi(t|x)\Theta(\xi-x),
\label{F_max}
\end{equation}
where $\Theta(x)$ is the Heaviside step function. By taking the derivative with respect to $\xi$ and recalling that 
$S_x(t|x)=0$ (survival is impossible starting at the boundary) we get the following expression for the probability density function (PDF) 
$\varphi_{\rm max}(\xi,t|x)$ of the maximum
\begin{equation}
\varphi_{\rm max}(\xi,t|x)=\frac{\partial S_\xi(t|x)}{\partial\xi}\Theta(\xi-x).
\label{pdf_max}
\end{equation}

Let us denote by $\Bigl\langle M(t) \bigl | x\Bigr\rangle$ the mean maximum value,
\begin{equation}
\Bigl\langle M(t) \bigl | x\Bigr\rangle=\int_{-\infty}^\infty\xi\varphi_{\rm max}(\xi,t|x) d\xi.
\label{def_mean_max}
\end{equation}
We have
\begin{equation}
\Bigl\langle M(t) \bigl | x\Bigr\rangle=\int_x^\infty\xi \frac{\partial S_\xi(t|x)}{\partial\xi}d\xi.
\label{mean_M1}
\end{equation}
At first sight this expression can be simplified by an integration by parts. This is, however, not possible because $S_\xi\rightarrow 1$ as 
$\xi\rightarrow\infty$ leading to a divergent result. The situation can be amended using $W_\xi$ instead of $S_\xi$. Substituting Eq. (\ref{s_w}) into Eq. (\ref{mean_M1}) followed by an integration by parts then yields
\begin{equation}
\Bigl\langle M(t) \bigl | x\Bigr\rangle=x+\int_x^\infty W_\xi(t|x) d\xi,
\label{mean_M2}
\end{equation}
where we have assumed that $W_\xi$ decreases faster than $1/\xi$ (i.e., $\xi W_\xi\rightarrow 0$ as $\xi\rightarrow\infty$). Attending that $W_\xi$ is always positive this equation shows, the otherwise obvious result, that the mean maximum is greater than the initial value. 

Following an analogous reasoning we can easily see that the moments of the maximum, defined by
\begin{equation}
\Bigl\langle M^n(t) \bigl | x\Bigr\rangle=\int_{-\infty}^\infty \xi^n \varphi_{\rm max}(\xi,t|x) d\xi,
\label{moments_max_0}
\end{equation}
are given by
\begin{equation}
\Bigl\langle M^n(t) \bigl | x\Bigr\rangle=x^n+n\int_x^\infty \xi^{n-1} W_\xi(t|x) d\xi,
\label{moments_max}
\end{equation}
$(n=1,2,3,\dots)$. In writing this equation we have assumed that $\xi^n W_\xi \to 0$ as $\xi\to\infty$ which is the condition imposed on $W_\xi$ for  moments to exist.

\subsection{The minimum}

We denote by
$$
m(t)=\min\{X(\tau); 0\leq\tau\leq t\}
$$
the minimum value attained by $X(t)$ during the time interval $(0,t)$, and let 
$$
\Phi_{\rm min}(\xi,t|x)= {\rm Prob}\{m(t)<\xi|X(0)=x\}
$$
be its distribution function.  Note that if $\xi<x$ the event $\{m(t)<\xi|X(0)=x\}$ implies that the process has crossed threshold $\xi$ at time $t$ or before. Hence the the distribution function agrees with the hitting probability to level $\xi$, {\it i.e.} $\Phi_{\rm min}(\xi,t|x)=W_\xi(t|x)$. On the other hand, when $\xi>x$ the event $\{m(t)<\xi|X(0)=x\}$ is certain and $\Phi_{\rm min}(\xi,t|x)=1$. Summing up
\begin{equation}
\Phi_{\rm min}(\xi,t|x)=\Theta(\xi-x)+W_\xi(t|x)\Theta(x-\xi).
\label{F_min}
\end{equation}
Let us denote by $\varphi_{\rm min}(\xi,t|x)$ the PDF of the minimum $m(t)$. Taking the derivative with respect to $\xi$ of $\Phi_{\rm min}$ and noting that $W_\xi(t|x)\delta(x-\xi)=\delta(x-\xi)$ (recall that $W_\xi(t|\xi)=1$) we get
\begin{equation}
\varphi_{\rm min}(\xi,t|x)=\frac{\partial W_\xi(t|x)}{\partial\xi}\Theta(x-\xi).
\label{pdf_min}
\end{equation}
The mean minimum value, defined as
\begin{equation}
\Bigl\langle m(t) \bigl | x\Bigr\rangle=\int_{-\infty}^\infty\xi\varphi_{\rm min}(\xi,t|x) d\xi,
\label{def_mean_min}
\end{equation}
is then given by
\begin{equation}
\Bigl\langle m(t) \bigl | x\Bigr\rangle=\int_{-\infty}^x\xi \frac{\partial W_\xi(t|x)}{\partial\xi}d\xi.
\label{mean_m1}
\end{equation}
An integration by parts yields 
$$
\Bigl\langle m(t) \bigl | x\Bigr\rangle=x-\lim_{\xi\rightarrow-\infty}[\xi W_{\xi}(t|x)]-\int_{-\infty}^x W_\xi(t|x) d\xi.
$$
Because $W_{-\infty}(t|x)=0$ ({\it i.e.,} hitting an infinite threshold is impossible) then, if we also assume that $W_\xi$ decreases faster than $1/|\xi|$, we have $\xi W_\xi\rightarrow 0$ as $\xi\rightarrow-\infty$ and 
\begin{equation}
\Bigl\langle m(t) \bigl | x\Bigr\rangle=x-\int_{-\infty}^x W_\xi(t|x) d\xi,
\label{mean_m2}
\end{equation}
which shows that the mean minimum value is indeed lower than the initial value.  

Analogously to the maximum value, the moments of the minimum are given by
\begin{equation}
\Bigl\langle m^n(t) \bigl | x\Bigr\rangle=x^n-n\int_{-\infty}^x \xi^{n-1} W_\xi(t|x) d\xi,
\label{moments_mean}
\end{equation}
as long as $W_\xi$ decreases faster than $|\xi|^{-n}$ as $\xi\to -\infty$. 

\section{Extremes and the escape problem}
\label{sec3}

The escape, or exit, problem addresses the question of whether or not a given process $X(t)$ starting inside an interval $(a,b)$ has  departed from it for the first time. The problem is solved when one knows the escape probability $W_{a,b}(t|x)$, which is defined as the probability of leaving $(a,b)$ at time $t$ (or before) for the first time and starting at $x\in (a,b)$. Closely related to the $W_{a,b}$ is the survival probability,
\begin{equation}
S_{a,b}(t|x)=1-W_{a,b}(t|x),
\label{S_ab}
\end{equation}
giving the probability that, starting inside $(a,b)$, the process has not exited this interval at time $t$ or before. 

For one dimensional diffusion processes, the escape probability satisfies the FPE \cite{gardiner,mp-feller}
\begin{equation}
\partial_t W_{a,b}(t|x)=f(x)\partial_x W_{a,b}(t|x)+\frac 12 D(x)\partial^2_{xx} W_{a,b}(t|x),
\label{fpe2}
\end{equation}
with initial and boundary conditions given by 
\begin{equation}
W_{a,b}(0|x)=0, \qquad  W_{a,b}(t|a)=W_{a,b}(t|a)=1.
\label{initial2}
\end{equation}

Note that $S_{a,b}(t|x)$ also obeys Eq. (\ref{fpe2}) but with initial and boundary conditions reversed; that is,
$$
S_{a,b}(0|x)=1, \qquad S_{a,b}(t|a)=S_{a,b}(t|a)=0.
$$

Extreme values related to the escape probability are essentially two: the maximum absolute value and the span. Let us next address them. 

\subsection{The maximum absolute value}

We now consider the maximum absolute value attained by $X(t)$ during the time span $(0,t)$. Denote by $G_{\rm max}(\xi,t|x)$ its distribution function,
\begin{equation}
G_{\rm max}(\xi,t|x)={\rm Prob}\left\{\max\bigl |X(\tau)\bigr|<\xi \bigl|X(0)=x\right\}, 
\label{abs_max_df}
\end{equation}
where $0\leq\tau\leq t$ and $\xi>0$. Certainly $\xi$ cannot be negative and hence 
$$
G_{\rm max}(\xi,t|x)=0, \qquad (\xi<0).
$$
In order to connect this distribution function with the escape problem we must distinguish two cases according to which the initial point is inside or outside the interval $(-\xi,\xi)$ spanned by the level $\xi>0$ of the absolute maximum. For the first case where  $-\xi<x<\xi$, we have
\begin{eqnarray*}
&&\Bigl\{\max\bigl |X(\tau)\bigr|<\xi; {0\leq\tau\leq t} \bigl|X(0)=x\Bigr\}\\ 
&&=
\Bigl\{-\xi<X(\tau)<\xi; 0\leq\tau\leq t \bigl |X(0)=x\Bigr\},
\end{eqnarray*}
meaning that during the time span $(0,t)$ the process $X(t)$ has not left the interval $(-\xi,\xi)$. Hence, the distribution function (\ref{abs_max_df}) coincides with the survival probability
$$
G_{\rm max}(\xi,t|x)=S_{-\xi,\xi}(t|x), \qquad (|x|<\xi).
$$

Note that when the initial value is outside the interval $(-\xi,\xi)$, the event $\{{\rm max} |X(\tau)|<\xi|X(0)=x\}$ ($0\leq\tau\leq t$) is impossible and 
$$
G_{\rm max}(\xi,t|x)=0, \qquad (|x|>\xi).
$$
Therefore, 
\begin{equation}
G_{\rm max}(\xi,t|x)=S_{-\xi,\xi}(t|x)\Theta(\xi-|x|),
\label{abs_max_prob}
\end{equation}
($\xi>0$). The PDF of the absolute maximum is defined by 
$$
g_{\rm max}(\xi,t|x)=\frac{\partial }{\partial \xi} G_{\rm max}(\xi,t|x).
$$

Substituting for Eq. (\ref{abs_max_prob}) and noting that 
$$
S_{-\xi,\xi}(t|x)\delta(\xi-|x|)=S_{-|x|,|x|}(t|x)\delta(\xi-|x|)=0,
$$
we get
\begin{equation}
g_{\rm max}(\xi,t|x)=\frac{\partial S_{-\xi,\xi}(t|x)}{\partial \xi}\Theta(\xi-|x|), 
\label{abs_max_PDF(a)}
\end{equation}
$(\xi>0).$ In terms of the escape probability $W_{-\xi,\xi}$ this PDF can be written as
\begin{equation}
g_{\rm max}(\xi,t|x)=-\frac{\partial W_{-\xi,\xi}(t|x)}{\partial \xi}\Theta(\xi-|x|). 
\label{abs_max_PDF(b)}
\end{equation}

Let us next evaluate the mean value of the absolute maximum defined by
$$
\Bigl\langle \max|X(t)| \bigl | x\Bigr\rangle=\int_0^\infty \xi g_{\rm max}(\xi,t|x) d\xi.
$$
From Eq. (\ref{abs_max_PDF(b)}) we have
$$
\Bigl\langle \max|X(t)| \bigl | x\Bigr\rangle=-\int_{|x|}^\infty \xi \frac{\partial W_{-\xi,\xi}(t|x)}{\partial \xi} d\xi.
$$
Integration by parts yields
\begin{equation}
\Bigl\langle \max|X(t)| \bigl | x\Bigr\rangle=|x|+\int_{|x|}^\infty W_{-\xi,\xi}(t|x)d\xi,
\label{average_abs_max}
\end{equation}
where we have taken into account that $W_{-|x|,|x|}(t|x)=1$ and made the reasonable assumption that the escape probability $W_{-\xi,\xi}$ decreases faster than $1/\xi$, that is, 
$\xi W_{-\xi,\xi}\rightarrow 0$ as $\xi\rightarrow\infty$.

Again, the moments of the maximum absolute value can be written as
\begin{equation}
\Bigl\langle \bigl(\max|X(t)|\bigr)^n \bigl | x\Bigr\rangle=|x|^n+n\int_{|x|}^\infty \xi^{n-1}W_{-\xi,\xi}(t|x)d\xi,
\label{moments_abs_max}
\end{equation}
$(n=1,2,3,\dots)$. These moments exist as long as $W_{-\xi,\xi}$ decreases faster than $|\xi|^{-n}$ as $|\xi|\to\infty$.

We finally remark that obtaining the minimum absolute value is meaningless, for this value is not a random variable: it is always zero.

\subsection{The range or span}

The range or span (also termed as ``the oscillation") of a random process $X(t)$ over the time interval $(0,t)$ is defined as the difference between the maximum and the minimum:
\begin{equation}
R(t)=M(t)-m(t).
\label{span1}
\end{equation}
This random quantity is either characterized by the distribution function,
$$
F_{R}(r,t|x)={\rm Prob}\{R(t)<r|X(0)=x\},
$$
or by the PDF 
\begin{equation}
f_R(r,t|x)=\frac{\partial}{\partial r} F_R(r,t|x).
\label{span_pdf1}
\end{equation}

We can relate the span distribution to the escape problem out of a variable interval. This connection is a bit convoluted and we show in 
Appendix \ref{span_pdf} that 
\begin{equation}
f_R(r,t|x)=\int_{x-r}^x\frac{\partial^2 S_{v,r+v}(t|x)}{\partial r^2} dv, 
\label{span_pdf3}
\end{equation}
$(r>0)$, where $S_{v,r+v}(t|x)$ is the survival probability in the (variable) interval $(v,r+v)$. 

Having the expression for the span PDF we next address the issue of the mean span:
\begin{equation}
\Bigl\langle R(t) \bigl | x\Bigr\rangle=\int_0^\infty r f_R(r,t|x) dr.
\label{E_R}
\end{equation}
Unfortunately the introduction of Eq. (\ref{span_pdf3}) into this definition leads to indeterminate boundary terms as the reader can easily check. In the Appendix \ref{average_span} we present a way of avoiding these inconsistencies and the final result reads
\begin{equation}
\Bigl\langle R(t) \bigl | x\Bigr\rangle=\int_{-\infty}^\infty\xi\frac{\partial S_\xi(t|x)}{\partial \xi} d\xi,
\label{E_R2}
\end{equation}
where $S_\xi(t|x)$ if the survival probability up to threshold $\xi$. Let us incidentally note the curious fact that the complete probability distribution of the span is determined by the escape problem out of the variable interval $(v,v+r)$ where $x-r<v<x$. However, the first moment of this distribution depends only on the first-passage problem of a varying threshold $-\infty<\xi<\infty$.  

In terms of the the hitting probability $W_\xi(t|x)$ the expression above for the mean span is greatly simplified. Indeed, substituting $S_\xi=1-W_\xi$ into Eq. (\ref{E_R2}), followed by an integration by parts, yield
\begin{eqnarray*}
\Bigl\langle R(t) \bigl | x\Bigr\rangle&=&-\int_{-\infty}^\infty\xi\frac{\partial W_\xi(t|x)}{\partial \xi} d\xi\\
&=&-\xi W_\xi(t|x)\biggr|_{\xi=-\infty}^{\xi=+\infty}+\int_{-\infty}^\infty W_\xi(t|x) d\xi.
\end{eqnarray*}
However, $W_\xi\rightarrow 0$ as $\xi\rightarrow\pm\infty$ ({\it i.e.,} crossing becomes impossible as threshold grows). If, in addition, we assume that this decay is faster than $1/\xi$, i.e., $\xi W_\xi\rightarrow 0$ ($\xi\rightarrow\pm\infty$), we have 
\begin{equation}
\Bigl\langle R(t) \bigl | x\Bigr\rangle=\int_{-\infty}^\infty W_\xi(t|x) d\xi.
\label{E_R3}
\end{equation}

It is worth noticing that one can arrive at this expression in a more direct way. In effect, recalling the definition of the range as the difference between the maximum and the minimum, we have
\begin{equation}
\Bigl\langle R(t) \bigl | x\Bigr\rangle=\Bigl\langle M(t) \bigl | x\Bigr\rangle-\Bigl\langle m(t) \bigl | x\Bigr\rangle,
\label{span_def}
\end{equation}
and substituting for Eqs. (\ref{mean_M2}) and (\ref{mean_m2}) we get 
$$
\Bigl\langle R(t) \bigl | x\Bigr\rangle=\int_{x}^\infty W_\xi(t|x) d\xi+\int_{-\infty}^x W_\xi(t|x) d\xi,
$$
which is Eq. (\ref{E_R3}).

There is no simple expressions,  beside Eq. (\ref{E_R3}), for the span higher moments as it is for the other extremes. In the present case moments have to be evaluated through their definition and the use of Eq. (\ref{span_pdf3})
\begin{eqnarray*}
\Bigl\langle R^n(t) \bigl | x\Bigr\rangle&=&\int_0^\infty r^n f_R(r,t|x) dr \\
&=&\int_0^\infty r^n dr \int_{x-r}^x \frac{\partial^2 S_{v,r+v}(t|x)}{\partial r^2} dv.
\end{eqnarray*}
This is quite unfortunate because the evaluation of span moments becomes a complicated business even numerically. The reason for not having a more convenient expression lies in the fact that maxima and minima are generally  correlated quantities and these correlations appear in all moments greater than the first one.

\section{The Wiener Process}
\label{sec4}

We now illustrate the expressions obtained above by reviewing one of the simplest, albeit very relevant, cases: the Wiener process or free Brownian motion, a diffusion process with zero drift and constant diffusion coefficient. Although some results related to first-passage and extremes for the Brownian motion can be traced as far back as to Bechelier,  Levy and Feller \cite{darling}, many results are found scattered in the mathematics and physics literature \cite{darling,blake}. It is, therefore, useful to have a summary of the main results about the extreme values of the Wiener process.  

\subsection{The maximum and the minimum}

The first-passage probability $W_c(t|x)$ to some threshold $x_c$ will be determined by the solution of the FPE (\ref{fpe})-(\ref{initial}) with $f(x)=0$ and $D(x)=D$. The time Laplace transform 
$$
\hat W_c(s|x)=\int_0^\infty e^{-st} W_c(t|x)dt
$$
leads to the following boundary-value problem
\begin{equation}
\frac{d^2\hat W_c}{dx^2}=(2/D)s\hat W_c, \qquad  \hat W_c(s|x_c)=1/s.
\label{fpe_wiener1}
\end{equation}
The solution to this problem that is finite for both $x>x_c$ and $x<x_c$ is straightforward and reads
$$
\hat W_c(s|x)=\frac 1s \exp\left\{-\sqrt{\frac{2s}{D}}\left|x-x_c\right|\right\}.
$$
Laplace inversion yields \cite{roberts}
\begin{equation}
W_c(t|x)={\rm Erfc}\left[\frac{|x-x_c|}{\sqrt{2Dt}}\right],
\label{wiener_Wc}
\end{equation}
where ${\rm Erfc}(z)$ is the complementary error function. The PDF of the maximum value is then given by Eq. (\ref{pdf_max}) or, equivalently, by
$$
\varphi_{\rm max}(\xi,t|x)=-\frac{\partial W_\xi(t|x)}{\partial\xi}\Theta(\xi-x),
$$
which results in the following truncated Gaussian density
\begin{equation}
\varphi_{\rm max}(\xi,t|x)=\left(\frac{2}{\pi Dt}\right)^{1/2}e^{-(\xi-x)^2/2Dt} \Theta(\xi-x).
\label{wiener_pdf_max}
\end{equation}
The mean maximum is then given by (cf Eqs. (\ref{def_mean_max}) or (\ref{mean_M2}))
\begin{equation}
\Bigl\langle M(t) \bigl | x\Bigr\rangle=x+\left(\frac{2Dt}{\pi}\right)^{1/2},
\label{wiener_max}
\end{equation}

Likewise, the PDF of the minimum value is given by (cf Eq. (\ref{pdf_min}))
\begin{equation}
\varphi_{\rm min}(\xi,t|x)=\left(\frac{2}{\pi Dt}\right)^{1/2}e^{-(x-\xi)^2/2Dt}\Theta(x-\xi),
\label{wiener_pdf_min}
\end{equation}
and the mean minimum reads
\begin{equation}
\Bigl\langle m(t) \bigl | x\Bigr\rangle=x-\left(\frac{2Dt}{\pi}\right)^{1/2}.
\label{wiener_min}
\end{equation}
Notice that both extreme values grow like $t^{1/2}$ as $t\rightarrow\infty$, the otherwise typical behavior of the Wiener process. 

These results can be generalized to include any moment of the maximum and the minimum.  By combining Eqs. (\ref{moments_max_0}) and (\ref{wiener_pdf_max}) we easily see that
\begin{eqnarray}
&&\Bigl\langle M^n(t) \bigl | x\Bigr\rangle=\nonumber \\ &&\frac{1}{\sqrt{\pi}}\sum_{k=0}^n \dbinom{n}{k}\Gamma\left(\frac{k+1}{2}\right)(2Dt)^{k/2}x^{n-k}
\label{moment_wiener_max}
\end{eqnarray} 
($n=1,2,3,\dots$). Following an analogous reasoning we show that the moments of of the minimum are
\begin{eqnarray}
&&\Bigl\langle m^n(t) \bigl | x\Bigr\rangle\nonumber \\ &&=\frac{1}{\sqrt{\pi}}\sum_{k=0}^n (-1)^k \dbinom{n}{k}\Gamma\left(\frac{k+1}{2}\right)(2Dt)^{k/2}x^{n-k}
\label{moment_wiener_min}
\end{eqnarray} 
($n=1,2,3,\dots$). With increasing $n$ these expressions become rather clumsy. We can get, however, simpler expressions if instead of the maximum or the minimum we consider their ``distance'' from the initial position. This is defined by $M(t)-x$ in the case of the maximum or by 
$x-m(t)$ for the minimum. We have
\begin{eqnarray}
\Bigl\langle \bigl(M(t)&-&x\bigr)^n(t)\bigl | x\Bigr\rangle=\Bigl\langle \bigl(x-m(t)\bigr)^n(t) \bigl | x\Bigr\rangle\nonumber \\
&=&\frac{1}{\sqrt{\pi}}\Gamma\left(\frac{n+1}{2}\right)(2Dt)^{n/2}.
\label{distance}
\end{eqnarray}
Both distances are equal showing the otherwise obvious symmetry of the process. 

\subsection{The maximum absolute value}

As shown in the previous section in order to characterize both the maximum absolute value and the span, we need to know the escape probability, $W_{a,b}(t|x)$, out of an interval $(a,b)$. For the maximum absolute value the interval is symmetric while for the span is asymmetric. 

The Laplace transform of the exit probability obeys the same equation than that of the first-passage probability,  Eq. (\ref{fpe_wiener1}), but with two boundary points:
$$
\hat W_{a,b}(s|a)=\hat W_{a,b}(s|b)=\frac 1s.
$$
The solution to this problem is
\begin{equation}
\hat W_{a,b}(t|x)=\frac{\cosh\sqrt{2s/D}[x-(a+b)/2]}{s\cosh\sqrt{2s/D}[(a-b)/2]}.
\label{escape_wiener1}
\end{equation}
The Laplace transform can be easily inverted \cite{roberts}. In the case of a symmetric interval ($-\xi,\xi$) the inverse transform is somewhat simpler yielding \cite{darling,roberts}
\begin{eqnarray}
W_{-\xi,\xi}(t|x)=1-\frac 2\pi\sum_{n=0}^\infty&&\frac{(-1)^n}{n+1/2}e^{-D(n+1/2)^2\pi^2t/\xi^2}\nonumber \\ 
&&\times \cos\bigl[(n+1/2)\pi x/\xi\bigr].
\label{escape_wiener2}
\end{eqnarray}

The PDF for the maximum absolute value, $g_{\rm max}(\xi,t|x)$, is readily obtained by introducing Eq. (\ref{escape_wiener2}) into Eq. (\ref{abs_max_PDF(b)}) (we will not write this expression).  Likewise the mean absolute maximum can be obtained from this form of the escape probability after substituting it into Eq. (\ref{average_abs_max}). The resulting expression is given by complicated infinite sums of exponential functions of little practical use, since from it is hard to figure out the asymptotic time behavior of that average.  It turns out to be more efficient to proceed from the Laplace transform of the average. We thus define
$$
\hat\mu(s|x)=\mathcal{L}\left\{\Bigl\langle \max|X(t)| \bigl | x\Bigr\rangle\right\}
$$
as the (time) Laplace transform of the mean absolute maximum. Transforming Eq. (\ref{average_abs_max}) yields
$$
\hat\mu(s|x)=\frac 1s |x|+\int_{|x|}^\infty \hat W_{-\xi,\xi}(s|x)d\xi.
$$
Plugging Eq. (\ref{escape_wiener1}) we see that the resulting integrals can be done in close form and  write
\begin{eqnarray}
\hat\mu(s|x)=\frac 1s |x|&+&\frac{\sqrt{2D}}{s^{3/2}}\cosh\Bigl(x\sqrt{2s/D}\Bigr)\nonumber \\ 
&\times& \left[\frac{\pi}{2}-\arctan e^{x\sqrt{2s/D}}\right]
\label{laplace_max_exact}
\end{eqnarray}
We now use this exact expression for the asymptotic analysis of the mean because, as Tauberian theorems prove \cite{tauberian}, the long time behavior of the mean is determined by the small $s$ behavior of its Laplace transform. It is a matter of simple algebra to show that as 
$s\rightarrow 0$ we have
$$
\hat \mu(s|x)=\frac 1s |x|+\frac{\pi}{4}\frac{\sqrt{2D}}{s^{3/2}}+O\left(\frac{1}{s^{1/2}}\right),
$$
which after Laplace inversion yields the asymptotic form of the mean absolute maximum
\begin{equation}
\Bigl\langle \max|X(t)| \bigl | x\Bigr\rangle \simeq |x|+\left(\frac{\pi Dt}{2}\right)^{1/2}+O\left(\frac{1}{t^{1/2}}\right),
\label{wiener_max_assym}
\end{equation}
showing again the $t^{1/2}$ growth.

\subsection{The span}

Let us finally describe the span of the Wiener process. As before we better work with Laplace transforms. Thus from 
Eq. (\ref{span_pdf3}) we write 
$$
\hat f_R(r,s|x)=-\frac{\partial^2 }{\partial r^2}\int_{x-r}^x \hat W_{v,r+v}(s|x) dv, 
$$
$(r>0)$, where the escape probability $\hat W_{v,r+v}(s|x)$ is given by Eq. (\ref{escape_wiener1}) (note that the second derivative can be pulled out of the integral because the lower limit is linear in $r$). 

For the Wiener process the escape probability is given by Eq.  (\ref{escape_wiener1}) and the integral above can be done in close form yielding
\begin{equation}
\hat f_R(r,s|x)=-(2D)^{1/2}\frac{\partial^2 }{\partial r^2}\left[\frac{1}{s^{3/2}}\tanh\left(\frac{s}{2D}\right)^{1/2}r\right].
\label{wiener_span_pdf}
\end{equation}
The Laplace transform of the mean span is then given by
\begin{eqnarray*}
&&\mathcal{L}\left\{\Bigl\langle R(t) \bigl | x\Bigr\rangle\right\}=\int_0^\infty r \hat f_R(r,s|x) dr = \\
&&-(2D)^{1/2} \int_0^\infty r \frac{\partial^2 }{\partial r^2}\left[\frac{1}{s^{3/2}}\tanh\left(\frac{s}{2D}\right)^{1/2}r\right] dr.
\end{eqnarray*}
Integration by parts yields
$$
\mathcal{L}\left\{\Bigl\langle R(t) \bigl | x\Bigr\rangle\right\}=\frac{(2D)^{1/2}}{s^{3/2}},
$$
and after inversion we get the exact result
\begin{equation}
\Bigl\langle R(t) \bigl | x\Bigr\rangle=2\left(\frac{2Dt}{\pi}\right)^{1/2},
\label{wiener_mean_span}
\end{equation}
which is, of course, the difference between the mean maximum (\ref{wiener_max}) and the mean minimum (\ref{wiener_min}) 
(see Eq. (\ref{span_def})). 

An interesting fact to note is that the long-time ratio between the mean absolute maximum  (\ref{wiener_max_assym}) and the mean span is fixed and given by
$$
\lim_{t\rightarrow\infty}\frac{\Bigl\langle \max|X(t)| \bigl | x\Bigr\rangle}{\Bigl\langle R(t) \bigl | x\Bigr\rangle}=\frac{\pi}{4},
$$
which means that at long times  the mean maximum absolute value is always smaller than the mean span. 

\section{Extremes of the Feller process}
\label{sec5}

The Feller process is another example  of diffusion process having linear drift and linear diffusion coefficient vanishing at the origin \cite{feller}. The process has been applied not only to the modeling of socio-economic systems (the CIR-Heston model \cite{cox}) but also in theoretical biology such as population dynamics and neuron firing processes \cite{ricciardi,gerstner}. It has been recently applied to reproduce cholera epidemics as a susceptible-infected-recovered model \cite{azaele}. It is also a significant model for single neuron dynamics where functionals of the first-passage time are employed to characterize the parameters of the model \cite{ditlevsen,bibbona}. 

The process is governed by a stochastic differential equation which in non-dimensional units (see \cite{mp-feller}) can be written as
\begin{equation}
dX(t)=-[ X(t)-\theta]dt+\sqrt{2X(t)}dW(t),
\label{feller0}
\end{equation}
where $W(t)$ is the Wiener process and $\theta>0$ is a dimensionless parameter --called saturation or normal level-- representing the value to which $X(t)$ is attracted. This parameter has a key role in the behavior of the process for it is related to the important question of the possibility of reaching the origin (which, for instance, in population dynamics would imply extinction \cite{capocelli}). Indeed, if $\theta\leq 1$ the probability of reaching the origin is greater than zero and $x=0$ is an accessible boundary. On the other hand, if $\theta>1$ such a probability is zero which renders the origin unaccessible (see \cite{mp-feller} for a simple proof and more details). 

The linear drift $f(x)=-(x-\theta)$ drives the process towards level $\theta$, a deterministic pull which is increased near the origin where the noise term is very small. In effect, the state-dependent diffusion coefficient $D(x)=2x$ for large values of $x$ enhances the the effect of noise while as $x$ goes to zero this effect vanishes. Therefore, when the process reaches the origin the drift drags it towards $\theta$ and since $\theta$ is positive the process remains always positive. The very fact that $X(t)$ never attains negative values makes the process a suitable candidate for modeling a number of phenomena in natural and social sciences \cite{mp-feller}. 

We now study the extreme values attained by the Feller process. We will basically obtain expressions for the maximum and minimum values because, due the positive character of the process, extremes such as the maximum absolute value coincide with the maximum. 

For $X(t)$ described by Eq. (\ref{feller0}) the first-passage probability to some threshold $\xi$ is the solution of the Fokker-Planck equation (cf. Eqs. (\ref{fpe})-(\ref{initial}))
\begin{equation}
\partial_t W_{\xi}(t|x)=-(x-\theta)\partial_x W_{\xi}(t|x)+x\partial^2_{xx} W_{\xi}(t|x),
\label{feller_fpe}
\end{equation}
with initial and boundary conditions given by
\begin{equation}
W_{\xi}(0|x)=0,  \qquad W_{\xi}(t|\xi)=1.
\label{feller_initial}
\end{equation}

We have recently proved that the solution to this problem for the time Laplace transform of $W_{\xi}$ is given by \cite{mp-feller} 
\begin{equation}
 \hat W_{\xi}(s|x)= \begin{cases} \frac{\displaystyle F(s,\theta,x)}{\displaystyle sF(s,\theta,\xi)}, & \quad \xi\geq x,  \smallskip \\
\frac{\displaystyle U(s,\theta,x)}{\displaystyle sU(s,\theta,\xi)}, & \quad \xi\leq x,
\end{cases}
\label{hitting_feller}
\end{equation}
where $F$ and $U$ are  confluent hypergeometric (Kummer) functions of first and second kind respectively \cite{mos}. 

\subsection{The maximum}

The distribution function of the maximum is related to the survival probability  $S_\xi(t|x)$ by Eq. (\ref{F_max}) which we write in terms of the hitting probability, $W_\xi(t|x)$, as
$$
\Phi_{\rm max}(\xi,t|x)=\left[1-W_\xi(t|x)\right]\Theta(\xi-x).
$$
In terms of $W_\xi$ the mean maximum is given by Eq. (\ref{mean_m2}):
$$
\Bigl\langle M(t) \bigl | x\Bigr\rangle=x+\int_x^{\infty} W_\xi(t|x) d\xi.
$$

Looking at Eq. (\ref{hitting_feller}) we see that for the Feller process the time Laplace transform of the distribution function and that of the mean are respectively given by
\begin{equation}
\hat\Phi_{\rm max}(\xi,s|x)=\frac 1s\left[1-\frac{F(s,\theta,x)}{F(s,\theta,\xi)}\right]\Theta(\xi-x).
\label{df_max_feller}
\end{equation}
and
\begin{equation}
\hat M(s|x)=\frac 1s\left[x+F(s,\theta,x)\int_x^{\infty}\frac{d\xi}{ F(s,\theta,\xi)} \right],
\label{max_feller1}
\end{equation}
where 
$$
\hat M(s|x)=\mathcal{L}\Bigl\{\Bigl\langle M(t) \bigl | x\Bigr\rangle\Bigr\}
$$
is the time Laplace transform of the mean maximum.

The PDF of the maximum is readily obtained by taking the derivative with respect to $\xi$ of the distribution function (\ref{df_max_feller}). We have
\begin{equation}
\hat\varphi_{\rm max}(\xi,s|x)=\frac{F(s,\theta,x)F'(s,\theta,\xi)}{sF^2(s,\theta,\xi)}\Theta(\xi-x),
\label{pdf_max_feller}
\end{equation}
where \cite{mos}
\begin{equation}
F'(s,\theta,\xi)=\frac{d}{d\xi}F(s,\theta,\xi)=\frac{s}{\theta}F(s+1,\theta+1,\xi).
\label{F'}
\end{equation}

Unfortunately the analytical inversion of these expressions to get their values in real time seems to be beyond reach, even though numerical inversion is always possible. We will find, nonetheless, some approximations that may be appropriate in practical cases.  

Let us first show that, like Brownian motion, the mean maximum value of the Feller process diverges as $t\to\infty$. One might have thought that since --unlike Brownian motion-- the Feller process possesses a force drifting the process towards the value $\theta$, the mean maximum would tend to a finite value (not far from $\theta$) as time increases. Let us show that this is not the case. Indeed, recalling the following property of the Laplace transform  \cite{tauberian}: 
\begin{equation}
\lim_{t\to\infty} f(t)=\lim_{s\to 0}\left[s\hat f(s)\right].
\label{s_limit}
\end{equation}
and the value of the Kummer function  $F(s=0,\theta,z)=1$ \cite{mos}, we see that the limit $s\to 0$ in (\ref{max_feller1}) leads to 
$$
\lim_{s\to 0}\left[s\hat M(s|x)\right]=x+\int_x^{\infty} d\xi=\infty.
$$
Whence
\begin{equation}
\Bigl\langle M(t) \bigl | x\Bigr\rangle\rightarrow \infty, \qquad (t \to \infty)
\label{max_inf}
\end{equation}
and the mean maximum diverges as time increases.

We next refine this asymptotic behavior. As is well known \cite{gardiner,redner1,weiss1}  the long-time expressions of first-passage probabilities are related to the mean first-passage time by (see also \cite{mp-feller}  for a simple derivation)
\begin{equation}
W_\xi(t|x)\simeq 1-e^{-t/T_\xi(x)}, \qquad (t\rightarrow\infty),
\label{asym1}
\end{equation}
where $T_\xi(x)$ is the mean first-passage time to threshold $\xi$ starting from $x$. Obviously this asymptotic expression is valid as long as the mean firs-passage time exists which is not always the case.  Thus, for instance, in the Wiener process $T_\xi(x)=\infty$ and the approximation given by Eq.  (\ref{asym1}) is meaningless.  For the Feller process this time exists and, as we have proved in \cite{mp-feller}, reads
\begin{equation}
T_\xi(x)= \begin{cases}(1/\theta)\int_x^\xi F(1,1+\theta,z)dz, & \quad \xi> x,  \medskip \\
 \int_\xi^x U(1,1+\theta,z)dz, & \quad \xi< x.
\end{cases}
\label{MFPT_feller}
\end{equation}

If the mean first-passage time exists, the distribution function of the maximum and its mean are, as $t\rightarrow\infty$, approximately given by 
\begin{equation}
\Phi_{\rm max}(\xi,t|x)\simeq e^{-t/T_\xi(x)}\Theta(\xi-x),
\label{asym2}
\end{equation}
and
\begin{equation}
\Bigl\langle M(t) \bigl | x\Bigr\rangle\simeq x+\int_x^{\infty}\left[1-e^{-t/T_\xi(x)}\right] d\xi,
\label{asym3}
\end{equation}
where here 
$$
T_\xi(x)=\frac 1\theta \int_x^\xi F(1,1+\theta,z)dz
$$
since the maximum is always greater than the initial point ($\xi> x$). Note that $1-e^{-t/T_\xi(x)}\rightarrow 0$ as $\xi\rightarrow\infty$ because the mean first-passage time to an infinite threshold is infinite and the integral in Eq. (\ref{asym3}) converges \footnote{In fact, for securing the convergence of the integral in Eq. (\ref{asym3}) we have to assume that $1-e^{-t/T_\xi(x)}\to 0$ faster than $\xi^{-1}$.}.

Equation (\ref{asym3}) is a compact expression that may be suitable for the numerical evaluation of the mean maximum for large values of time. As far as I can see it is, however, of little use for further analytical approximations.

Let us thus obtain another asymptotic expansion of the maximum value which is valid for large values of the initial position $x$.  Our starting point is the time Laplace transform of the mean maximum given by Eq. (\ref{max_feller1}). Assume now that $x\rightarrow\infty$ we can then use the following approximation \cite{mos}
\begin{equation}
F(s,\theta,x)=\frac{\Gamma(\theta)}{\Gamma(s)}e^xx^{s-\theta}\left[1+O\left(x^{-1}\right)\right],
\label{asym_kumer_F}
\end{equation}
and since $\xi>x$ then $\xi$ is also large and we have an analogous expression for $F(s,\theta,\xi)$. Substituting both approximations into Eq. (\ref{max_feller1}) we get as $x\rightarrow\infty$ 
$$
\hat M(s|x)\simeq \frac 1s\left[x+e^xx^{s-\theta}\int_x^{\infty}e^{-\xi}\xi^{\theta-s}d\xi\right].
$$
But the integral can written in terms of the incomplete Gamma function $\Gamma(1+\theta-s,x)$ and within the same approximation we have \cite{mos}
$$
\int_x^{\infty}e^{-\xi}\xi^{\theta-s}d\xi=\Gamma(1+\theta-s,x)\simeq e^{-x}x^{\theta-s}\left[1+O\left(x^{-1}\right)\right].
$$
Substituting into the previous equation yields $\hat M(s|x)\simeq(x+1)/s+O(x^{-1})$ which after Laplace inversion results in the simple asymptotic approximation:
\begin{equation}
\Bigl\langle M(t) \bigl | x\Bigr\rangle\simeq x+1+O\left(x^{-1}\right).
\label{asym4}
\end{equation}
Despite its appeal, this approximations merely means that the mean maximum value grows at the same pace as it does the starting value, as can be otherwise seen from Eq. (\ref{mean_M2}).

\subsection{The minimum}

We recall from Sec. \ref{sec2} that in terms of the hitting probability the distribution function of the minimum is (see Eq. (\ref{F_min}))
$$
\Phi_{\rm min}(\xi,t|x)=\Theta(\xi-x)+W_\xi(t|x)\Theta(x-\xi).
$$
The mean minimum is given in Eq. (\ref{mean_m2}) where, due to the positive character of the Feller process, we replace $-\infty$ in the lower limit of integration by $0$: 
\begin{equation}
\Bigl\langle m(t) \bigl | x\Bigr\rangle=x-\int_{0}^x W_\xi(t|x) d\xi.
\label{min_feller0}
\end{equation}

Taking into account Eq. (\ref{hitting_feller}), the time Laplace transform of these quantities reads
\begin{equation}
\hat\Phi_{\rm min}(\xi,s|x)=\frac 1s\left[\Theta(\xi-x)+\frac{U(s,\theta,x)}{U(s,\theta,\xi)}\Theta(x-\xi)\right],
\label{df_min_feller}
\end{equation}
and
\begin{equation}
\hat m(s|x)=\frac 1s\left[x-U(s,\theta,x)\int_0^{x}\frac{d\xi}{ U(s,\theta,\xi)} \right],
\label{min_feller1}
\end{equation} 
where $\hat m(s|x)$ is the time Laplace transform of the mean minimum and the $U$'s are Kummer functions of second kind \cite{mos}.  

Taking the $\xi$-derivative of Eq. (\ref{df_min_feller}) we get the PDF of the minimum
\begin{equation}
\hat\varphi_{\rm min}(\xi,s|x)=-\frac{U(s,\theta,x)U'(s,\theta,\xi)}{sU^2(s,\theta,\xi)}\Theta(x-\xi),
\label{pdf_min_feller}
\end{equation}
where \cite{mos}
\begin{equation}
U'(s,\theta,\xi)=\frac{d}{d\xi}U(s,\theta,\xi)=-s U(s+1,\theta+1,\xi).
\label{U'}
\end{equation}

Starting form Eq. (\ref{min_feller1}) and using the property given in Eq. (\ref{s_limit}) we can obtain the limiting value of the mean minimum when $t\to\infty$. We begin with the relationship between Kummer functions $U$ and $F$ \cite{mos}:
\begin{eqnarray}
U(s,\theta,x)&=&\frac{\Gamma(1-\theta)}{\Gamma(1+s-\theta)} F(s,\theta, x) \label{U_F}\\ 
&+&\frac{\Gamma(\theta-1)}{\Gamma(s)}x^{1-\theta} F(1+s-\theta,2-\theta, x).
\nonumber 
\end{eqnarray}
Recalling that as $s\to 0$ $F(s,\theta,x)\to 1$ and $\Gamma(s)\to\infty$ we see that $U(s,\theta,x)\to 1$. Hence
$$
\lim_{s\to 0}\left[s\hat m(s|x)\right]=x-\int_0^{x}d\xi=0,
$$
and from Eq. (\ref{s_limit}) we conclude that
\begin{equation}
\Bigl\langle m(t) \bigl | x\Bigr\rangle\to 0, \qquad (t\to\infty). 
\label{min_inf}
\end{equation}
The mean minimum thus converges to the origin as time increases. 
 
We next refine this crude estimate for large, but finite, values of time. When $t\to\infty$ and after using the asymptotic form of the hitting probability given in Eq. (\ref{asym1}), we get
\begin{equation}
\Phi_{\rm min}(\xi,t|x)\simeq 1-e^{-t/T_\xi(x)} \Theta(x-\xi),  
\label{min_asym1}
\end{equation}
($t\to\infty$), where $T_\xi(x)$ is the MFPT to threshold $\xi$ which when $\xi< x$ is given by (cf. Eq. (\ref{MFPT_feller}))
$$
T_\xi(x)=\int_\xi^x U(1,1+\theta,z)dz, \qquad (\xi< x).
$$

Substituting Eq. (\ref{asym1}) into Eq. (\ref{min_feller0}) we find the following long-time approximation of the mean minimum
\begin{equation}
\Bigl\langle m(t) \bigl | x\Bigr\rangle\simeq \int_{0}^x e^{-t/T_\xi(x)} d\xi, \qquad(t\to\infty).
\label{min_asym2}
\end{equation}

Likewise the long-time behavior of the maximum value discussed above, these asymptotic expressions related to the minimum value are more appropriate for numerical evaluation rather than for obtaining further practical analytical approximations. 

We will find, nonetheless, approximations of the mean minimum when the initial value $x$ is small and close to the origin.  
Our starting point is the expression of the Laplace transform of the mean minimum given in Eq. (\ref{min_feller1}). We next assume that $x$ is small then from Eq, (\ref{U_F}) and the fact that $F(a,b,x)=1+O(x)$ \cite{mos} we write
\begin{eqnarray}
U(s,\theta,x)&=&\frac{\Gamma(1-\theta)}{\Gamma(s+1-\theta)}\bigl[1+O(x)\bigr] \nonumber \\
&+&  \frac{\Gamma(\theta-1)}{\Gamma(s)}x^{1-\theta}\bigl[1+O(x)\bigr].
\label{u_expan0}
\end{eqnarray}
Note that the leading term in this expansion depends on whether $\theta>1$ or $\theta<1$. We, therefore, distinguish the cases:

(i) $\theta>1$ (recall that in this case the origin is unattainable by the dynamical evolution of the process \cite{mp-feller}). Now Eq. (\ref{u_expan0}) yields the approximation
\begin{equation}
U(s,\theta,x) \simeq \frac{\Gamma(\theta-1)}{\Gamma(s)}x^{1-\theta}\bigl[1+O(x)\bigr].
\label{u_expan1}
\end{equation}
Since the integral in Eq. (\ref{min_feller1}) runs from $\xi=0$ to $\xi=x$ when $x$ is small $\xi$ is also small. We can thus use approximation (\ref{u_expan1}) for $U(s,\theta,\xi)$ inside the integral and write
\begin{eqnarray*}
\int_0^{x}\frac{d\xi}{ U(s,\theta,\xi)} &\simeq& \frac{\Gamma(s)}{\Gamma(\theta-1)}\int_0^x \xi^{\theta-1}d\xi \\ 
&=&\frac{\Gamma(s)}{\Gamma(\theta-1)}\frac{x^\theta}{\theta}.
\end{eqnarray*}
Plugging this approximation along with Eq. (\ref{u_expan1}) into  Eq. (\ref{min_feller1})  we get $\hat m(s|x)\simeq x(1-1/\theta)/s$ which after Laplace inversion yields
\begin{equation}
\Bigl\langle m(t) \bigl | x\Bigr\rangle\simeq \left(1-\frac 1\theta\right)x, \qquad (x\to 0).
\label{min_asym3}
\end{equation}

(ii) $\theta<1$ (the origin is attainable \cite{mp-feller}). In this case Eq. (\ref{u_expan0}) provides  the following consistent expansion
\begin{equation}
U(s,\theta,x)=\frac{\Gamma(1-\theta)}{\Gamma(1+s-\theta)}+
\frac{\Gamma(\theta-1)}{\Gamma(s)} x^{1-\theta}+O(x).
\label{u_expan2}
\end{equation}
Substituting this into the integral in Eq. (\ref{min_feller1}), expanding the denominator to the lowest order in $\xi$ (recall that $\xi< x$ is small when $x$ is small) and integrating we obtain
\begin{equation}
\int_0^x \frac{d\xi}{U(s,\theta,\xi)}=\frac{\Gamma(1+s-\theta)}{\Gamma(1-\theta)} x+O(x^{2-\theta}).
\label{int_1/u}
\end{equation}
 
In order to proceed further it is more convenient to use an integral representation for the Kummer function $U(s,\theta,x)$ (which multiplies the integral in Eq. (\ref{min_feller1})) instead of using the expansion (\ref{u_expan2}). Thus, taking into account the transformation formula \cite{mos}
$$
U(s,\theta,x)=x^{1-\theta} U(s+1-\theta,2-\theta,x),
$$
and using the integral representation \cite{mos}
$$
U(a,b,x)=\frac{1}{\Gamma(a)}\int_0^\infty e^{-xz}z^{a-1}(1+z)^{b-a-1} dz,
$$
we get
\begin{equation}
U(s,\theta,x)=\frac{x^{1-\theta}}{\Gamma(1+s-\theta)}\int_0^\infty e^{-xz} z^{-\theta}\left(\frac{z}{1+z}\right)^s dz.
\label{u_expan3}
\end{equation}

Substituting Eqs. (\ref{int_1/u}) and (\ref{u_expan3}) into Eq. (\ref{min_feller1}) results in the following approximate expression for the Laplace transform of the mean minimum
\begin{eqnarray}
\hat m(s|x)&=&\frac 1s\Biggl[x-\frac{x^{2-\theta}}{\Gamma(1-\theta)} \\ \nonumber
&\times& \int_0^\infty e^{-xz} z^{-\theta} \left(\frac{z}{1+z}\right)^s dz\Biggr]
+O(x^{3-2\theta}).
\label{min_4}
\end{eqnarray}
In the Appendix \ref{min_feller} we invert this equation and obtain the power law
\begin{equation}
\Bigl\langle m(t) \bigl | x\Bigr\rangle \simeq A(t) x^{2-\theta}, \qquad  (x\to 0), 
\label{min_asym_final}
\end{equation}
where
\begin{equation}
A(t)=\frac{1}{\Gamma(2-\theta)}\left(\frac{e^{-t}}{1-e^{-t}}\right)^{1-\theta}.
\label{a}
\end{equation} 

We finally note that $\theta<1$ implies $2-\theta>1$ and the mean minimum (\ref{min_asym_final}) decays sharper than the linear law (\ref{min_asym3}), the latter applicable when $\theta>1$. This is a somewhat intuitive and interesting behavior meaning that as the process starts near the origin the average minimum tends faster to $x=0$ if the boundary is accessible than otherwise.

\subsection{The span}

As shown in Sec. \ref{sec3} the PDF of the range or span is given by Eq. (\ref{span_pdf3}) which in terms of the escape probability and taking the Laplace transform with respect to time reads
\begin{equation}
\hat f_R(r,s|x)=-\int_{x-r}^x\frac{\partial^2\hat W_{v,r+v}(s|x)}{\partial r^2} dv.
\label{span_pdf_feller}
\end{equation}
We have proved elsewhere \cite{mp-feller} that in the Feller process the Laplace transform of the escape probability is given by
\begin{widetext} 
\begin{equation}
\hat W_{v,v+r}(s|x)=\frac{\bigl[U(s,\theta,v+r)-U(s,\theta,v)\bigr]F(s,\theta,x)-\bigl[F(s,\theta,v+r)-F(s,\theta,v)\bigr]U(s,\theta,x)}
{s\bigl[F(s,\theta,v)U(s,\theta,v+r)-F(s,\theta,v+r)U(s,\theta,v)\bigr]}.
\label{exit_prob}
\end{equation}
\end{widetext}
Unfortunately the introduction of Eq. (\ref{exit_prob}) into Eq. (\ref{span_pdf_feller}) does not lead to an expression amenable to further analytical simplifications, being only suitable for numerical work. 

The mean span is simpler because we only need to know the hitting probability instead of the escape probability. Thus substituting Eq. (\ref{hitting_feller}) into the Laplace transform of Eq. (\ref{E_R3}) we get
\begin{eqnarray}
\hat R(s|x)=&&\frac 1s \Biggl[U(s,\theta,x)\int_0^x \frac{d\xi}{U(s,\theta,\xi)} \nonumber \\
&&+F(s,\theta,x)\int_x^\infty \frac{d\xi}{F(s,\theta,\xi)}\Biggr],
\label{feller_span}
\end{eqnarray}
where $\hat R(s|x)$ is the Laplace transform of the mean span,
$$
\hat R(s|x)=\int_0^\infty e^{-st} \Bigl\langle R(t)\bigl|x\Bigr\rangle dt.
$$ 

Note that the analytical simplifications carried out for the maximum and the minimum are of no use here, for when $x$ is small we can obtain a simpler expression for the first integral but not for the second, while when $x$ is large the situation is reversed. A similar difficulty arises when $t\to \infty$. We, therefore, conclude that Eq. (\ref{feller_span}) seems to be only appropriate for numerical work.

\section{Summary of main results and closing remarks}
\label{sec6}

We have reviewed the relationship between level-crossing problems and the distribution of extreme values for continuous-time random processes. We have compiled and rederived in a simpler way many general results which would remain otherwise scattered in the literature. We have applied them to the Wiener and Feller processes; the latter, we believe, for the first time.    

Let us recall that level-crossing problems are solved when one knows the hitting probability (in first-passage problems) or the exit probability (in escape problems). We have denoted these probabilities by $W_\xi(t|x)$ and $W_{a,b}( t|x)$ respectively. In both cases $x$ is the initial value of the process whereas $\xi$ is the threshold, or critical value, and $(a,b)$ is the exit interval. For one-dimensional diffusion processes characterized by drift $f(x)$ and diffusion coefficient $D(x)$ both probabilities satisfy the FPE
$$
\partial_t W(t|x)=f(x)\partial_x W(t|x)+\frac 12 D(x)\partial^2_{xx} W(t|x)
$$
with initial condition $W(0|x)=0$. The boundary conditions are $W_\xi(t|\xi)=1$ (first-passage) or $W_{a,b}(t|a)=W_{a,b}(t|b)=1$ (escape). 

We denote by $M(t|x)$ and $m(t|x)$ the maximum and minimum values attained by the process during the time span $(0,t)$ and starting at $x$ at $t=0$. The PDF's of these random quantities are respectively given by 
$$
\varphi_{\rm max}(\xi,t|x)=-\frac{\partial W_\xi(t|x)}{\partial\xi}\Theta(\xi-x),
$$
and
$$
\varphi_{\rm min}(\xi,t|x)=\frac{\partial W_\xi(t|x)}{\partial\xi}\Theta(x-\xi), 
$$
where $\varphi_{\rm max}(\xi,t|x)d\xi={\rm Prob}\{\xi<M(t)<\xi+d\xi|x\}$ and similarly for $\varphi_{\rm min}(\xi,t|x)$. 

Moments of order $n=1,2,3,\dots$ of the maximum and the minimum are also written in terms of the hitting probability as 
$$
\Bigl\langle M^n(t) \bigl | x\Bigr\rangle=x^n+n\int_x^\infty \xi^{n-1}W_\xi(t|x) d\xi,
$$
and
$$
\Bigl\langle m^n(t) \bigl | x\Bigr\rangle=x^n-n\int_0^x \xi^{n-1}W_\xi(t|x) d\xi.
$$

If we denote by $g_{\rm max}(\xi,t|x)$ the PDF of the maximum absolute value of the random process $X(t)$, {\it i.e.} 
$$
g_{\rm max}(\xi,t|x)d\xi={\rm Prob}\Bigl\{\xi<\max|X(t)|<\xi+d\xi\bigl|x\Bigr\},
$$
then 
$$
g_{\rm max}(\xi,t|x)=-\frac{\partial W_{-\xi,\xi}(t|x)}{\partial\xi}\Theta(\xi-|x|),
$$
where $W_{-\xi,\xi}(t|x)$ is the escape probability out of the symmetric interval $(-\xi,\xi)$. Moments of this statistic are
$$
\Bigl\langle \bigl(\max|X(t)|\bigr)^n \bigl | x\Bigr\rangle=|x|^n+n\int_{|x|}^\infty \xi^{n-1}W_{-\xi,\xi}(t|x)d\xi.
$$
 
The second quantity related to the escape problem is the range or span, that is, the difference between maximum and minimum $R(t)=M(t)-m(t)$. We define the PDF of this random oscillation as 
$$
f_R(r,t|x)dr={\rm Prob}\Bigl\{r<R(t)<r+dt\bigl|x\Bigr\}, 
$$
$(r>0)$,  and it reads:  
$$
f_R(r,t|x)=-\int_{x-r}^x\frac{\partial^2 W_{v,r+v}(t|x)}{\partial r^2} dv,
$$
where $W_{v,r+v}(t|x)$ is the escape probability out of the variable interval $(v,v+r)$,  where $v$ runs from $x-r$ to $x$.  The mean range has a simple expression in terms of the hitting probability $W_\xi(t|x)$ to a variable threshold:
$$
\Bigl\langle R(t) \bigl | x\Bigr\rangle=\int_{-\infty}^{\infty} W_\xi(t|x) d\xi.
$$
Due to correlations between maximum and minimum, the moments of the span have no simple expression in terms of the hitting probability and we need to know the entire escape probability to evaluate moments higher than the first (see the end of Sec. \ref{sec3}). 

We have applied the above results to the Wiener process. The PDF's of the maximum and minimum are given by simple truncated Gaussian densities and the PDF's of the maximum absolute values and of the span are given by more complicated expressions written in terms of infinite series. We refer the reader to Sec. \ref{sec4} for the explicit expressions of these quantities and more information about mean values and  moments. 

We have finally  dealt with the maximum and minimum values achieved by the Feller process. This is a linear diffusion process which never  attains negative values. The behavior of the process near the origin is governed by a dimensionless parameter $\theta>0$ (cf. Eq. (\ref{feller0})). When $\theta<1$ the origin is an accessible boundary while if $\theta>1$ it is unattainable \cite{mp-feller}.

In a recent work we solved the level-crossing problem for the Feller process and obtained the time Laplace transform of the hitting and escape probabilities \cite{mp-feller} (see Eqs. (\ref{hitting_feller}) and (\ref{exit_prob}) respectively). The PDF's of the maximum and the minimum are respectively given by (see Eqs. (\ref{pdf_max_feller})-(\ref{F'}) and 
Eqs. (\ref{pdf_min_feller})-(\ref{U'}))
$$
\hat\varphi_{\rm max}(\xi,s|x)=\frac{F(s+1,\theta+1,\xi)}{\theta F^2(s,\theta,\xi)}F(s,\theta,x)\Theta(\xi-x)
$$
and
$$
\hat\varphi_{\rm min}(\xi,s|x)=\frac{U(s+1,\theta+1,\xi)}{U^2(s,\theta,\xi)}U(s,\theta,x)\Theta(x-\xi),
$$
where $F(a,b,z)$ and $U(a,b,z)$ are Kummer functions \cite{mos} and
$$
\hat\varphi(\xi,s|x)=\int_0^\infty e^{-st} \varphi(\xi,t|x) dt
$$
is the time Laplace transform of $\varphi$.

These exact expressions for the Laplace transform of the PDF's do not seem to be invertible analytically. However, as we have shown in  Sec. \ref{sec5} there exist asymptotic analytical approximations in real time. Thus, as $t\to\infty$ and after taking the derivative with respect to $\xi$ of Eqs. (\ref{asym2}) and (\ref{min_asym1}), we have
$$
\varphi_{\rm max}(\xi,t|x)\simeq t\frac{F(1,1+\theta,\xi)}{\theta T^2_\xi(x)}e^{-t/T_\xi(x)}\Theta(\xi-x)
$$
and 
$$
\varphi_{\rm min}(\xi,t|x)\simeq t\frac{U(1,1+\theta,\xi)}{T^2_\xi(x)}e^{-t/T_\xi(x)}\Theta(x-\xi),
$$
where $T_\xi(x)$ is the mean first-passage time given in Eq. (\ref{MFPT_feller}). 

The Laplace transforms of the mean maximum and minimum are
$$
\hat{M}(s|x)=\frac 1s\left[1+F(s,\theta,x)\int_x^\infty \frac{d\xi}{F(s,\theta,\xi)}\right],
$$
and 
$$
\hat{m}(s|x)=\frac 1s\left[1-U(s,\theta,x)\int_0^x \frac{d\xi}{U(s,\theta,\xi)}\right].
$$
As $t\to\infty$ these mean values in real time are approximated by
$$
\Bigl\langle M(t) \bigl | x\Bigr\rangle\simeq x + \int_x^\infty\left[1-e^{-t/T_\xi(x)}\right]d\xi,
$$
and
$$
\Bigl\langle m(t) \bigl | x\Bigr\rangle\simeq  \int_0^x e^{-t/T_\xi(x)} d\xi,
$$
where $T_\xi(x)$ is given in Eq. (\ref{MFPT_feller}).  We have also proved that as $t\to\infty$, the mean maximum diverges while the mean minimum converges towards the origin:
$$
\lim_{t\to\infty}\Bigl\langle M(t) \bigl | x\Bigr\rangle=\infty, \qquad \lim_{t\to\infty}\Bigl\langle m(t) \bigl | x\Bigr\rangle=0.
$$

An interesting behavior is provided by the mean minimum as $x\to 0$. Here we find a different result according to whether the natural boundary $x=0$ is unaccessible ($\theta>1$) or accessible ($\theta<1$) by the dynamics of the process. In the first case the average minimum decays linearly with $x$ while in the second it decays by a steeper power law. This is summarized by ($x\to 0$)
$$
\Bigl\langle m(t) \bigl | x\Bigr\rangle\simeq \begin{cases} (1-1/\theta)x, & \quad \theta>1,  \smallskip \\
A(t) x^{2-\theta}, & \quad \theta<1,
\end{cases}
$$ 
where $A(t)$ is defined in Eq. (\ref{a}).

\medskip
In this paper we have studied the extreme problem in a complete fashion where all extreme statistics are assumed to depend on the initial value $X(0)=x$ taken by the process under study. However, in many practical situations  and in some theoretical settings it is not possible to know the exact value of the initial value and one has to resort to averaging over all possible values of $x$.  In such cases one can, for instance, define the averaged (or reduced) maximum PDF as
\cite{katja}
$$
\varphi_{\rm max}(\xi,t)=\int_{-\infty}^\infty \varphi_{\rm max}(\xi,t|x)p(x)dx,
$$
where $p(x)$  is the PDF of the initial value. In those cases where the underlying process $X(t)$ is stationary it is sensible to assume that the process has been functioning since the infinitely distant past so that the initial PDF $p(x)$ is given by the stationary distribution:
$$
p(x)=\lim_{t_0\to-\infty}p(x,t=0|x_0,t_0),
$$
where $p(x,t|x_0,t_0)$ is the propagator of the underlying process. Obviously such a procedure requires the existence of a stationary distribution, something that, for instance, the Wiener process does not possess but Feller process does ({\it i.e.,} the Gamma distribution \cite{mp-feller}). This averaging procedure and some practical applications of the formalism are under present investigation.

\acknowledgments

Partial financial support from the Ministerio de Ciencia e Innovaci\'{o}n under Contract No. FIS 2009-09689 is acknowledged.

\appendix

\section{The probability distribution of the span}
\label{span_pdf}

Let us denote by $F_2(\xi,\eta,t|x)$ the joint distribution function of the maximum and the minimum:
$$
F_2(\xi,\eta,t|x)={\rm Prob}\{M(t)<\xi, m(t)<\eta|X(0)=x\}.
$$
Note that the event $\{M(t)<\xi\}$ is the union of two disjoint events:
\begin{eqnarray*}
\{M(t)<\xi\}&=&\{M(t)<\xi,m(t)<\eta\} \\
&\cup&\{M(t)<\xi, m(t)>\eta\},
\end{eqnarray*}
where we have dropped the dependence on the initial value $x$ which is, nonetheless, implied in all what follows. We thus have
\begin{eqnarray*}
&&{\rm Prob}\{M(t)<\xi, m(t)<\eta\} \\ 
&&={\rm Prob}\{M(t)<\xi\} -{\rm Prob}\{M(t)<\xi, m(t)>\eta\},
\end{eqnarray*}
but (see Eqs. (\ref{F_max_def}) and (\ref{F_max}))
$$
{\rm Prob}\{M(t)<\xi\}=S_\xi(t|x)\Theta(\xi-x),
$$
where $S_\xi(t|x)$ is the survival probability up to the single threshold $\xi$. If, on the other hand, $S_{\eta,\xi}(t|x)$ is the survival probability 
of the interval $(\eta,\xi)$ one easily realizes that
$$
{\rm Prob}\{M(t)<\xi, m(t)>\eta\}=S_{\eta,\xi}(t|x)\Theta(\xi-x)\Theta(x-\eta).
$$
Collecting results we write
$$
F_2(\xi,\eta,t|x)=S_\xi(t|x)\Theta(\xi-x)-S_{\eta,\xi}(t|x)\Theta(\xi-x)\Theta(x-\eta).
$$

The joint PDF of the maximum and the minimum, defined as the second derivative of the joint distribution function
$$
f_2(\xi,\eta,t|x)=\frac{\partial^2}{\partial\xi\partial\eta}F_2(\xi,\eta,t|x),
$$
is then given by 
\begin{eqnarray*}
f_2(\xi,\eta,t|x)&=&-\frac{\partial}{\partial\xi}\Biggl[\frac{\partial S_{\eta,\xi}}{\partial\eta}\Theta(\xi-x)\Theta(x-\eta) \\
&-&S_{\eta,\xi}(t|x)\delta(x-\eta)\Theta(\xi-x)\Biggr].
\end{eqnarray*}
Recalling that starting at any boundary point renders survival impossible we see that 
$$
S_{\eta,\xi}(t|x)\delta(x-\eta)=S_{x,\xi}(t|x)\delta(x-\eta)=0.
$$
Hence
\begin{eqnarray*}
f_2(\xi,\eta,t|x)&=&-\frac{\partial^2 S_{\eta,\xi}}{\partial\xi\partial\eta}\Theta(\xi-x)\Theta(x-\eta) \\
&-&\frac{\partial S_{\eta,\xi}}{\partial\eta}\delta(\xi-x)\Theta(x-\eta),
\end{eqnarray*}
but again $S_{\eta,x}(t|x)=0$, so that 
$$
\frac{\partial S_{\eta,\xi}}{\partial\eta}\delta(\xi-x)=\frac{\partial}{\partial\eta}\bigl[S_{\eta,x}(t|x)\delta(\xi-x)\bigr]=0.
$$
Therefore
\begin{equation}
f_2(\xi,\eta,t|x)=-\frac{\partial^2 S_{\eta,\xi}}{\partial\xi\partial\eta}\Theta(\xi-x)\Theta(x-\eta).
\label{f_2}
\end{equation}

In terms of the joint density the PDF of the span, Eq. (\ref{span_pdf1}), is given by 
\begin{equation}
f_R(r,t|x)=\int_{-\infty}^\infty d\xi \int_{-\infty}^\infty \delta[r-(\xi-\eta)]f_2(\xi,\eta,t|x) d\eta,
\label{span_pdf0}
\end{equation}
which, after substituting for Eq. (\ref{f_2}) and integrating the delta function, yields
\begin{equation}
f_R(r,t|x)=-\int_{x-r}^x\left.\frac{\partial^2 S_{\eta,\xi}(t|x)}{\partial\eta\partial\xi}\right|_{\xi=r+\eta} d\eta, 
\label{span_pdf2}
\end{equation}
where $r>0$ (recall that, by definition, $R(t)$ is always positive). This expression for $f_R$ is more conveniently written by making the change of variables 
$$
r=\xi-\eta, \quad v=\eta.
$$
Indeed, $d\eta=dv$ and 
$$
\left.\frac{\partial^2 S_{\eta,\xi}}{\partial\eta\partial\xi}\right|_{\xi=r+\eta}= -\frac{\partial^2 S_{v,r+v}}{\partial r^2}+
\frac{\partial^2 S_{v,r+v}}{\partial v \partial r}.
$$
Substituting into Eq. (\ref{span_pdf2}) and taking into account (recall that $S_{x,x+r}(t|x)=S_{x-r,x}(t|x)=0$) 
\begin{eqnarray*}
&&\int_{x-r}^x \frac{\partial^2 S_{v,r+v}(t|x)}{\partial r \partial v} dv=\frac{\partial}{\partial r}\int_{x-r}^x \frac{\partial S_{v,r+v}(t|x)}{\partial v}\\
&&= \frac{\partial}{\partial r}\Bigl[S_{x,x+r}(t|x)- S_{x-r,x}(t|x)\Bigr]=0,
\end{eqnarray*}
we finally get
\begin{equation}
f_R(r,t|x)=\int_{x-r}^x\frac{\partial^2 S_{v,r+v}(t|x)}{\partial r^2} dv,
\label{span_pdf3a}
\end{equation}
$(r>0)$, which is Eq. (\ref{span_pdf3}).

\section{The mean span}
\label{average_span}

In order to avoid divergencies appearing in the evaluation of the mean span we proceed as follows. Instead of using Eq. (\ref{span_pdf3}) as the expression for the span PDF  we will use the following expression of $f_R$ which results of combining Eqs. (\ref{f_2}) and (\ref{span_pdf0}):
\begin{eqnarray*}
&&f_R(r,t|x)=\\ 
&&-\int_{-\infty}^\infty d\xi \int_{-\infty}^\infty d\eta\frac{\partial^2 S_{\eta,\xi}}{\partial\xi\partial\eta} \delta[r-(\xi-\eta)]\Theta(\xi-x)\Theta(x-\eta).
\end{eqnarray*}
Plugging into
$$
\Bigl\langle R(t) \bigl | x\Bigr\rangle=\int_0^\infty rf_R(r,t|x)dr,
$$
and performing the integration over $r$ using the delta function we obtain
\begin{eqnarray}
&&\Bigl\langle R(t) \bigl | x\Bigr\rangle= \label{E_R1} \\ 
&&-\int_{-\infty}^\infty d\xi \int_{-\infty}^\infty d\eta (\xi-\eta)\frac{\partial^2 S_{\eta,\xi}}{\partial\xi\partial\eta}\Theta(\xi-x)\Theta(x-\eta). \nonumber
\end{eqnarray}
We rewrite this equation as
\begin{eqnarray}
&&\Bigl\langle R(t) \bigl | x\Bigr\rangle= \label{a1} \\
&&-\int_{-\infty}^\infty d\xi \Theta(\xi-x) \xi \frac{\partial}{\partial\xi}\int_{-\infty}^\infty d\eta \Theta(x-\eta)
\frac{\partial S_{\eta,\xi}}{\partial\eta} \nonumber \\ 
&&+\int_{-\infty}^\infty d\eta \Theta(x-\eta) \eta \frac{\partial}{\partial\eta}\int_{-\infty}^\infty d\xi \Theta(\xi-x)
\frac{\partial S_{\eta,\xi}}{\partial\xi}, \nonumber
\end{eqnarray}
but
\begin{eqnarray*}
\int_{-\infty}^\infty \Theta(x&-&\eta)\frac{\partial S_{\eta,\xi}}{\partial\eta} d\eta =\int_{-\infty}^x \frac{\partial S_{\eta,\xi}}{\partial\eta} d\eta \\
&=& S_{\xi,x}(t|x)-S_{-\infty,\xi}(t|x).
\end{eqnarray*}
However, $S_{\xi,x}(t|x)=0$ and
$$
S_{-\infty,\xi}(t|x)=S_{\xi}(t|x),
$$
because the escape problem out of the semi-infinite interval $(-\infty,\xi)$ coincides with the first-passage problem to threshold $\xi$. Hence
\begin{equation}
\int_{-\infty}^\infty \Theta(x-\eta)\frac{\partial S_{\eta,\xi}}{\partial\eta} d\eta=-S_\xi(t|x).
\label{a2}
\end{equation}
Proceeding similarly we get
\begin{equation}
\int_{-\infty}^\infty \Theta(\xi-x)\frac{\partial S_{\eta,\xi}}{\partial\xi} d\xi=S_\eta(t|x).
\label{a3}
\end{equation}
Plugging Eqs. (\ref{a2})-(\ref{a3}) into Eq. (\ref{a1}) and applying the Heaviside functions $\Theta(\xi-x)$ and $\Theta(x-\eta)$ we get
$$
\Bigl\langle R(t) \bigl | x\Bigr\rangle=\int_x^\infty \xi\frac{\partial S_\xi(t|x)}{\partial\xi} d\xi+\int_{-\infty}^x \eta\frac{\partial S_\eta(t|x)}{\partial\eta} d\eta.
$$
That is 
$$
\Bigl\langle R(t) \bigl | x\Bigr\rangle=\int_{-\infty}^\infty\xi\frac{\partial S_\xi(t|x)}{\partial \xi} d\xi,
$$
which is Eq. (\ref{E_R2}).

\section{Derivation of Eq. (\ref{min_asym_final})}
\label{min_feller}

We write the Laplace inversion of Eq. (80) in the form
\begin{eqnarray}
\Bigl\langle m(t) \bigl | x\Bigr\rangle&=&x-\frac{x^{2-\theta}}{\Gamma(1-\theta)}  \nonumber \\
&\times& \int_0^\infty e^{-xz} z^{-\theta} \mathcal{L}^{-1}\left\{\frac 1s\left(\frac{z}{1+z}\right)^s\right\} dz  \nonumber \\
&+&O(x^{3-2\theta}),
\label{c1}
\end{eqnarray}
where $\mathcal{L}^{-1}\{\cdot\}$ stands for Laplace inversion. Noting that 
$$
\left(\frac{z}{1+z}\right)^s=\exp\left[s\ln\left(\frac{z}{1+z}\right)\right],
$$
and using \cite{roberts}
$$
\mathcal{L}^{-1}\left\{\frac{e^{-as}}{s}\right\}=\Theta(t-a),
$$
where $\Theta(\cdot)$ is the Heaviside step function, we have
\begin{eqnarray*}
\mathcal{L}^{-1}\left\{\frac 1s\left(\frac{z}{1+z}\right)^s\right\}&=&\Theta\left[t+\ln\left(\frac{z}{1+z}\right)\right] \\
&=&\Theta\left(z-\frac{e^{-t}}{1-e^{-t}}\right).
\end{eqnarray*}
Hence
\begin{eqnarray*}
&&\int_0^\infty e^{-xz} z^{-\theta} \mathcal{L}^{-1}\left\{\frac 1s\left(\frac{z}{1+z}\right)^s\right\} dz \\
&&=\int_{\frac{e^{-t}}{1-e^{-t}}}^\infty e^{-xz}z^{-\theta}dz=x^{\theta-1}\Gamma\left(1-\theta,\frac{xe^{-t}}{1-e^{-t}}\right),
\end{eqnarray*}
where $\Gamma(a,z)$ is the incomplete Gamma function \cite{mos}. Substituting into Eq. (\ref{c1}) yields
\begin{eqnarray*}
\Bigl\langle m(t) \bigl | x\Bigr\rangle=x&-&\frac{x}{\Gamma(1-\theta)}\Gamma\biggl(1-\theta, \frac{xe^{-t}}{1-e^{-t}}\biggr) \\
&+&O(x^{3-2\theta}).
\end{eqnarray*}
For small values of $x$ and $t>0$ the argument of the incomplete Gamma function is small and we can use the following expansion \cite{mos}
$$
\Gamma(a,z)=\Gamma(a)-\frac{z^{a}}{a}+O(z^{a+1}),
$$
with the result
$$
\Bigl\langle m(t) \bigl | x\Bigr\rangle=\frac{1}{\Gamma(2-\theta)}\left(\frac{e^{-t}}{1-e^{-t}}\right)^{1-\theta} x^{2-\theta}+O(x^{3-2\theta}),
$$
which is Eq. (\ref{min_asym_final}).


\begin{thebibliography}{99}

\bibitem{gardiner} C. W. Gardiner, {\it Handbook of Stochastic Methods} (Springer, Berlin, 1985). 
\bibitem{redner1} S. Redner, {\it A Guide to First-Passage Processes} (Cambridge University Press, Cambridge, England, 2001).
\bibitem{weiss1} G. H. Weiss, {\it First-Passage Time Problems in Chemical Physics}, in Advances in Chemical Physiscs, Vol. 13, edited by I. Prigogine (J. Wiley, Hoboken, NJ, 2007).
\bibitem{katja} K. Lindenberg and B. West, J. Stat. Phys. {\bf 42}, 201 (1986).
\bibitem{mporra} J. Masoliver and J. Porr\`a, Phys. Rev. Lett. {\bf 75}, 189 (1995).
\bibitem{west} B. J. West, Chemical Physics {\bf 284}, 45 (2002).
\bibitem {eichner} J. Eichner, J. Kantelhardt, A. Bunde and S. Havlin Phys. Rev. E {\bf 73}, 016130 (2006).
\bibitem{shlesinger} M. F. Shlesinger, Nature (London) {\bf 450}, 40 (2007).
\bibitem{condamin} S. Condamin, O. B\'enichou, V. Tejedor, R. Volturiez and J. Klafter, Nature {\bf 450}, 77 (2007).
\bibitem{salvadori} G. Salvadori, S, De Michelle, N, T. Kottegoda, and R. Rosso, {\it Extremes in Nature} (Springer, Berlin, 2007).
\bibitem{maso_pere_07} J. Masoliver and J. Perell\'o, Phys. Rev. E {\bf 75}, 046110 (2007).
\bibitem{mont_maso} M. Montero and J. Masoliver, Eur. Phys. J. B {\bf 57}, 181 (2007).
\bibitem{maso_pere_08} J. Masoliver and J. Perell\'o, Phys. Rev. E {\bf 78}, 056104 (2008).
\bibitem{maso_pere_09} J. Masoliver and J. Perell\'o, Phys. Rev. E {\bf 80}, 016108 (2009).
\bibitem{mario} J. Perell\'o, M. Guti\'errez-Roig and J. Masoliver Phys. Rev. E {\bf 84}, 066110 (2011).
\bibitem{gumbel} E. J. Gumbel, {\it Statistics of Extremes} (Dover, New York, 2004).
\bibitem{leadbetter} M. R. Leadbetter, G. Lindgren and H. Rootzen, {\it Extremes and Related Properties of Random Sequences and Processes} (Springer, Berlin, 2011).
\bibitem{darling} A. J. F. Siegert, Phys. Rev {\bf 81}, 617 (1951); D. A. Darling and A. F. J. Siegert, Ann. Math. Statist. {\bf 24}, 624 (1953).
\bibitem{blake} I. F. Blake and W. C. Lindsay, IEEE Transactions on Information Theory, {\bf IT-19}, 295 (1973).
\bibitem{berman} S. M. Berman, {\it Sojourns and Extremes of Stochastic Processes} (Wadsworth and Brooks/Cole, Belmont, CA, 1992). 
\bibitem{yuste} E. Abad, S. B. Yuste and K. Lindenberg, Phys. Rev. E {\bf 86}, 061120 (2012).
\bibitem{mp-feller} J. Masoliver and J. Perell\'o, Phys. Rev. E {\bf 86}, 041116 (2012).
\bibitem{roberts}  G. E. Roberts and H. Kaufman, {\it Table of Laplace Transforms} (W. B. Sauders, Philadelphia, 1966).
\bibitem{tauberian} R. A. Handelsman and J. S. Lew, SIAM J.  Math.  Analysis {\bf 5}, 425-451 (1974).
\bibitem{feller} W. Feller, Ann. Math. {\bf 54}, 173-182 (1951); Ann. Math. {\bf 55}, 468-519 (1952); Trans. Am. Math. Soc. {\bf 71}, 1--31 (1954).
\bibitem{cox} J. C. Cox, J. E. Ingersoll and S. A. Ross, Econometrica {\bf 53}, 385 (1985).
\bibitem{ricciardi} R.M. Capocelli and L.M. Ricciardi, J. Theor. Biol. {\bf 40}, 369--387 (1973).
\bibitem{gerstner} W. Gerstner and W. M. Kistler, {\it Spiking Neuron Models} (Cambridge University Press, Cambridge, 2002).
\bibitem{azaele} S. Azaele, A. Maritan, E. Bertuzzo, I. Rodriguez-Iturbe and A. Rinaldo, Phys. Rev. E 81, 051901 (2010).
\bibitem{ditlevsen} S. Ditlevsen and P. Lansky, Phys. Rev. E 73, 061910 (2006).
\bibitem{bibbona} E. Bibbona, P. Lansky and R. Sirovich, Phys. Rev. E 81, 031916 (2010).
\bibitem{capocelli} R.M. Capocelli and L.M. Ricciardi, Theoretical Population Biology {\bf 5}, 28--41 (1974).
\bibitem{mos} W. Magnus, F. Oberhettinger and R. P. Soni, {\it Formulas and Theorems for the Special Functions of Mathematical Physics} (Springer-Verlag, Berlin and New York, 1966).


\end{thebibliography}
\end{document}